\documentclass[twocolumn,nofootinbib,
showpacs,prl,aps,floatfix]{revtex4}

\usepackage{graphicx}
\def\la{\langle}
\def\ra{\rangle}
\newcommand {\fexp} [1] {\exp \left( #1 \right)}
\newcommand {\fabsq}[1] {\left| #1 \right|^2}
\newcommand {\Msi}{\times 10^6\, \mbox{s}^{-1}}

\newcommand {\si} {\, \mbox{s}^{-1}}
\newcommand {\mum}{\, \mu \mbox{m}}
\newcommand {\cms}{\, \mbox{cm/s}}

\begin{document}
\title{An atom diode}
\author{A. Ruschhaupt}
\affiliation{Departamento de Qu\'\i mica-F\'\i sica,
Universidad del Pa\'\i s Vasco, Apdo. 644, Bilbao, Spain}
\author{J. G. Muga}
\affiliation{Departamento de Qu\'\i mica-F\'\i sica,
Universidad del Pa\'\i s Vasco, Apdo. 644, Bilbao, Spain} 


\begin{abstract}
An atom diode, i.e.,
a device that lets the ground state atom pass in one direction but not 
in the opposite direction in a velocity range is devised. It is       
based on the adiabatic transfer achieved with two lasers and a 
third laser potential that reflects the ground state. 
\end{abstract}
\maketitle

The detailed control of internal and/or translational atomic states 
is a major goal 
of quantum optics.   
Optical elements in which the roles of light and matter are reversed 
such as mirrors, gratings, 
interferometers,  
or beam splitters made of laser light or magnetic fields 
allow to 
manipulate atomic waves. 
Further handling of the atoms 
is inspired in 
electronic devices and integrated circuits: atom chips
\cite{folman.2000} and atom-optic circuits \cite{schneble.2003}
have been realized recently.  
The aim of this letter is to propose simple models  
for an ``atom diode'', a device built with laser light 
that lets the neutral atom in its ground state pass in one direction but not 
in the opposite direction for a range of incident velocities.   
A diode is a very basic circuit element and many applications are 
feasible in 
atomic trapping, or logic gates 
for quantum information processing.   
 
More specifically our goal is to model an atom-field interaction 
so that the ground state atom is transmitted when
traveling, say,  from left
to right, and it is reflected if coming from the right. 
We shall describe (effective) three-level and two-level atom models, for 
simplicity in one dimension,   
to achieve the desired behaviour. In both cases 
the atom is in an excited state after
being transmitted and, in principle, excited atoms could cross 
the diode ``backwards'', i.e., from right to left. 
Nevertheless, an irreversible decay 
from the excited state to the ground
state, would effectively block any backward motion.  

\begin{figure}[t]
\begin{center}
\includegraphics[angle=0,width=0.48\linewidth]{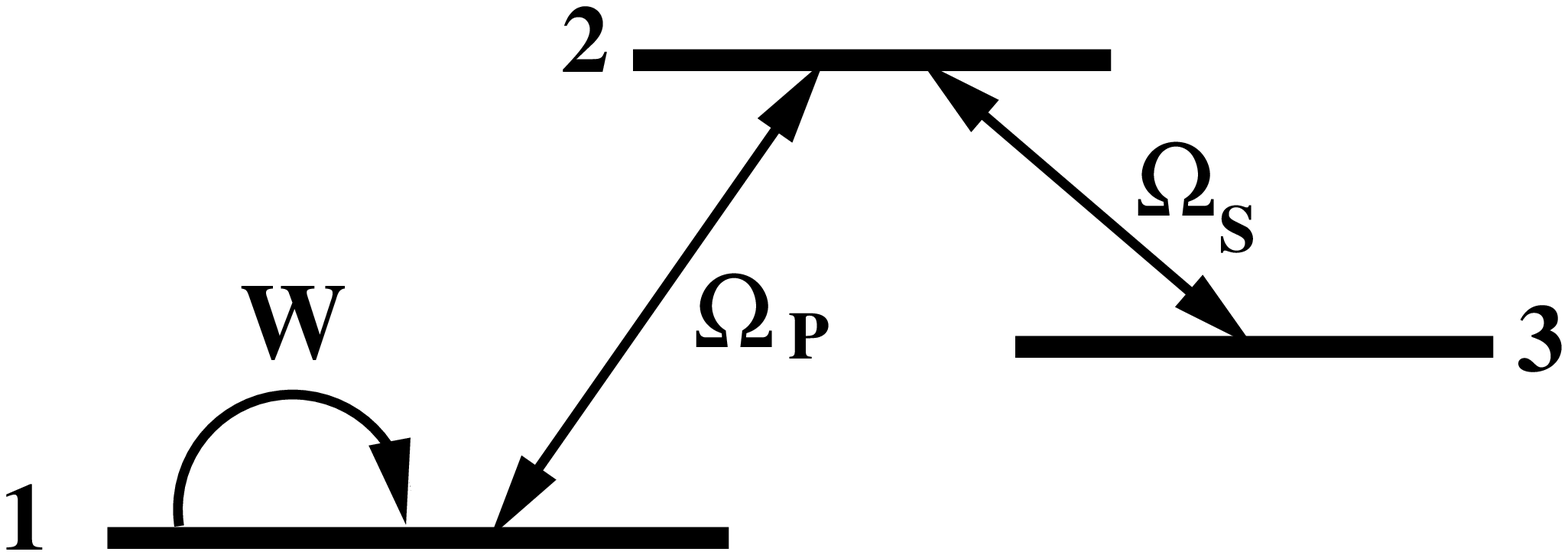}\hspace{0.2cm}
\includegraphics[angle=0,width=0.48\linewidth]{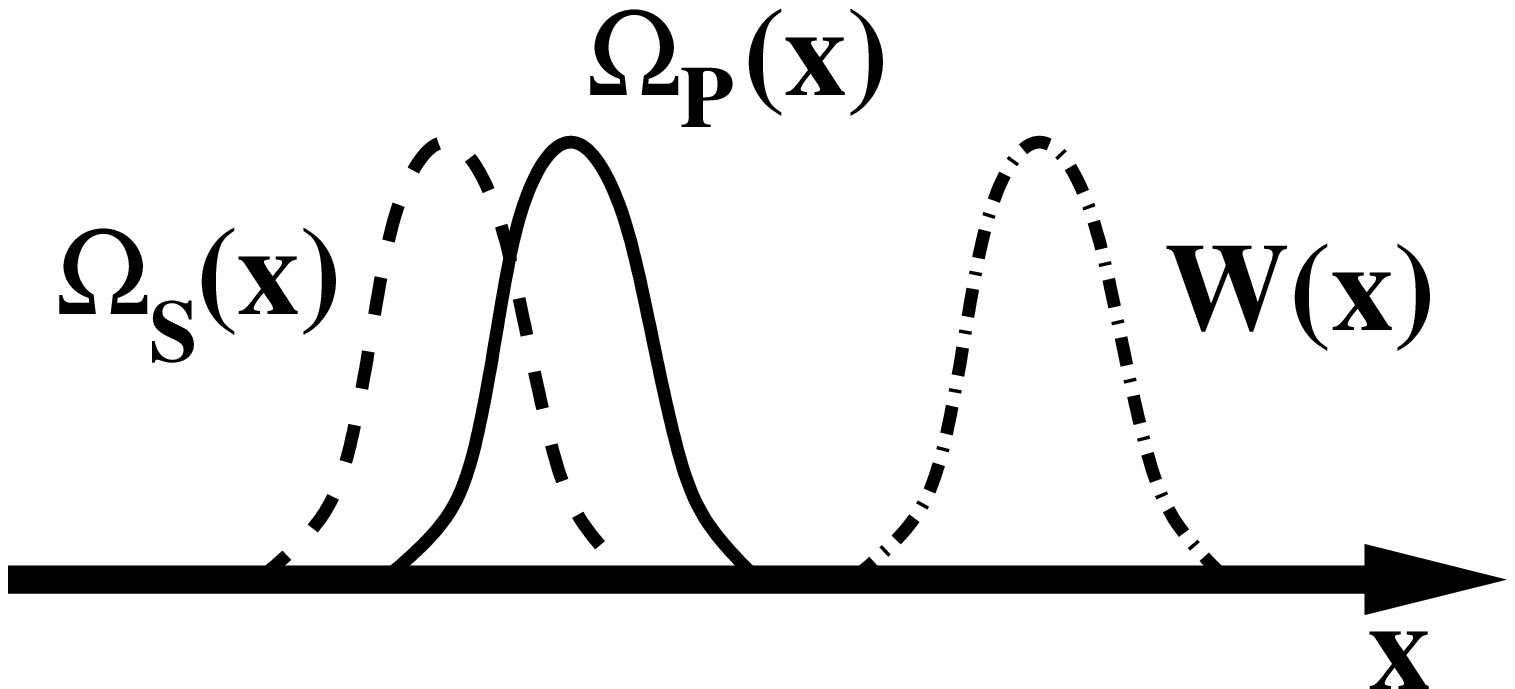}
\end{center}
\caption{\label{fig1}Schematic connection of the atom levels
by the different lasers (left figure) and 
location of the different lasers (right figure).}
\end{figure}

Let us denote by $R^l_{\beta\alpha} (v)$ ($R^r_{\beta\alpha} (v)$)
the scattering amplitudes for incidence with (modulus of) velocity $v$
from the left (right) in channel $\alpha$ and reflection in channel $\beta$.
Similarly we denote 
by $T^l_{\beta\alpha} (v)$ ($T^r_{\beta\alpha} (v)$) the scattering 
amplitude for incidence in channel $\alpha$ with velocity  $v$ from
the left (right) and transmission in channel $\beta$
to the right (left).
We define, for incidence in the ground state,
\begin{eqnarray*}
\hat{R} (v) =\left\{\begin{array}{ccc}
R^l_{11} (v) & : & v > 0\\
R^r_{11} (-v) & : & v < 0
\end{array}\right.,
\hat{T} (v) =\left\{\begin{array}{ccc}
T^l_{31} (v) & : & v > 0\\
T^r_{31} (-v) & : & v < 0
\end{array}\right.
\end{eqnarray*}
The potential will be such that $\fabsq{\hat{T} (v)} \approx 1$, 
$\fabsq{\hat{R} (v)} \approx 0$ and
$\fabsq{\hat{T}(-v)} \approx 0$, $\fabsq{\hat{R} (-v)} \approx 1$
($v > 0$).
The basic idea is to combine two lasers that achieve STIRAP (stimulated  
Raman adiabatic passage) with an additional
reflecting interaction for the ground state.
The STIRAP method is well known 
\cite{bergmann.1998}) and consists of an adiabatic transfer of population 
between levels 1 and 3 by two partially overlapping (in time or space)
laser beams, see Fig. \ref{fig1}.  
The pump laser 
couples the atomic levels 1 and 2 with  Rabi frequency $\Omega_P$, 
and the Stokes laser couples the states 2 and 3
with Rabi frequency $\Omega_S$.
We assume here that these two lasers are on resonance with the corresponding 
transitions. 
We shall need in addition a third laser causing an 
effective reflecting potential $V$ 
for the ground state component. 
It could be realized by an intense laser with a large 
positive detuning $\Delta$ (laser frequency minus the transition frequency) 
with respect to a transition 
with a fourth level,
$V(x)= W(x)\hbar/2=\Omega_{14}(x)^2\hbar/4\Delta$, 
$\Omega_{14}$ being the corresponding Rabi frequency.
Due to the large detuning, there is no pumping so that this type of 
coupling has a purely 
mechanical effect. 
Neglecting decay, the resulting Hamiltonian for the atomic state,
within the rotating wave approximation, and 
in the appropriate interaction picture to get rid of any time dependence, is
\begin{eqnarray}
H_{3L} = \frac{p_x^2}{2m} + \frac{\hbar}{2} \left(\begin{array}{ccc}
W(x) & \Omega_P(x) & 0\\
\Omega_P (x) & 0 & \Omega_S (x)\\
0 & \Omega_S (x) & 0
\end{array}\right),
\label{ham}
\end{eqnarray}
where $p_x=i\hbar\frac{\partial}{\partial x}$ is the momentum operator.
%
%
The shapes of the Rabi frequencies and the reflecting potential 
in the model are Gaussian, 
$\Omega_P (x)= \hat{\Omega} \;\Pi(x, x_P)$,
$\Omega_S (x)= \hat{\Omega} \;\Pi(x, x_S)$,
$W (x) = \hat{W} \;\Pi(x, x_W)$
with
\begin{eqnarray*}
\Pi (x, x_0)&=& \fexp{-\frac{(x-x_0)^2}{2 \Delta x^2}},
\end{eqnarray*}
but similar shapes do not alter the results in any significant way. 
We shall also assume for simplicity that the shapes and widths of pump laser,
Stokes laser and
additional  potential are equal. 
The location of the three laser beams  
is shown in Fig. \ref{fig1}.

If the atom is incident from the left in the ground state, it will be 
transfered by
STIRAP to the third state 
so it is not affected by $V(x)$,
and will be transmitted, i.e. the transmission probability 
$\fabsq{\hat{T} (v)} \approx 1$,
while the other reflection and transmission probabilities
for left incidence in the first state will be approximately zero. 
If the atom is incident from the right in the ground state, it
is reflected by the (high enough) potential $V$.
Therefore $\fabsq{\hat{T} (-v)} \approx 0
\neq  \fabsq{\hat{T} (v)}$ and
$\fabsq{\hat{R} (-v)} \approx 1$ ($v>0$).
The other reflection and transmission probabilities will be 
also approximately
zero.

This behavior is indeed observed solving numerically the
stationary Schr\"odinger equation with  
Eq. (\ref{ham}) 
by  
the
invariant imbedding method \cite{singer.1982,band.1994}.
%
\begin{figure}[t]
\begin{center}
\includegraphics[angle=-90,width=0.95\linewidth]{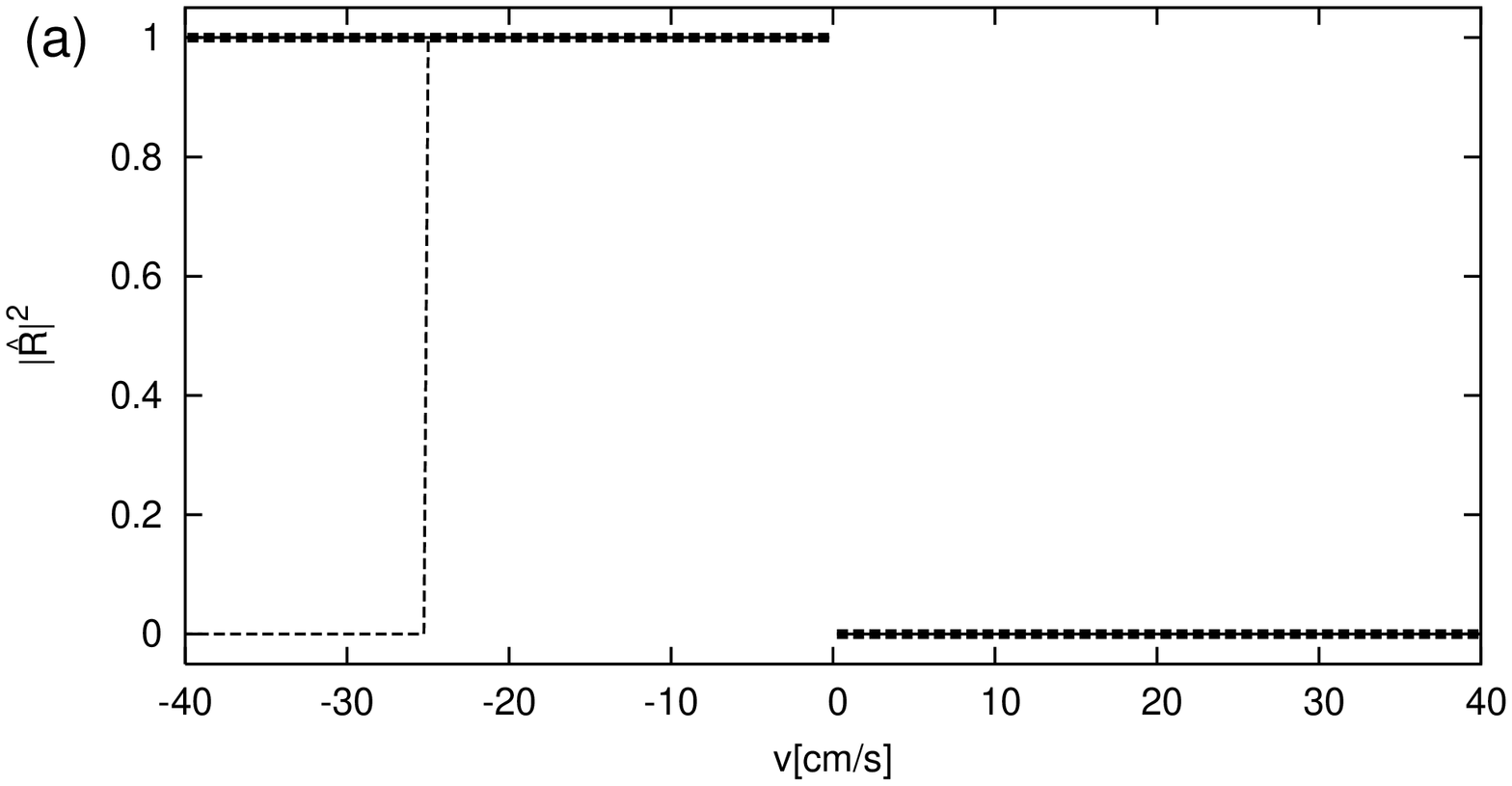}

\includegraphics[angle=-90,width=0.95\linewidth]{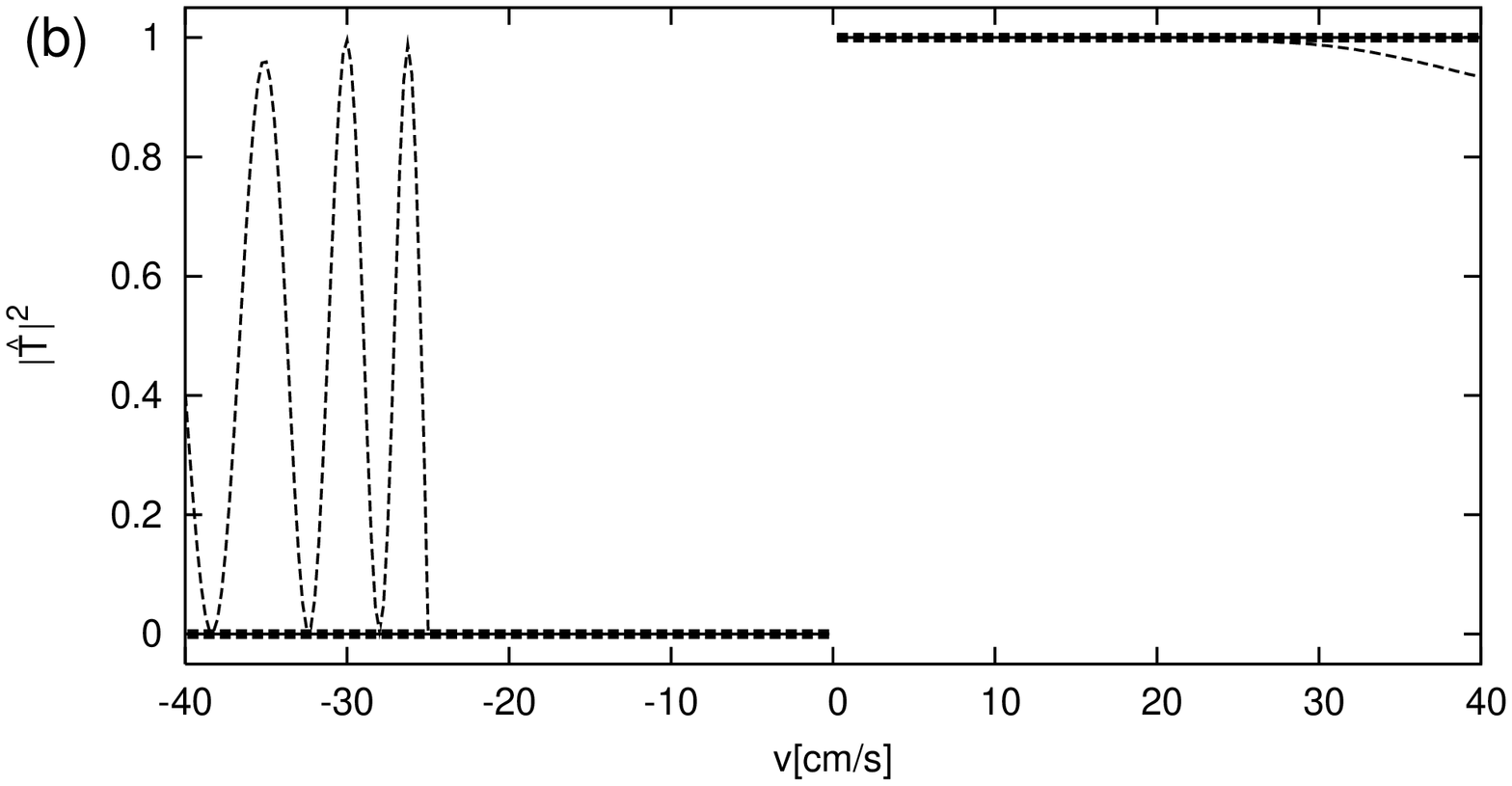}
\end{center}
\caption{\label{fig2}(a) Reflection probability $\fabsq{\hat{R}(v)}$ and
(b) transmission probability $\fabsq{\hat{T} (v)}$;
the mass is the mass of Neon, $\Delta x = 15 \mum$,
$x_S = 140 \mum$, $x_P = 170 \mum$;
three level atom: $x_W = 260 \mum$,
$\hat{\Omega}= 0.2 \Msi$, $\hat{W}=20 \Msi$ (thin dashed line),
$\hat{\Omega}= 1 \Msi$, $\hat{W}= 100 \Msi$ (thick dashed line);
two level atom: $\hat{f}^2 = 100 \Msi$ (solid line, coincides with thick
dashed line).}
\end{figure}
%
%
\begin{figure}[t]
\begin{center}
\includegraphics[angle=-90,width=\linewidth]{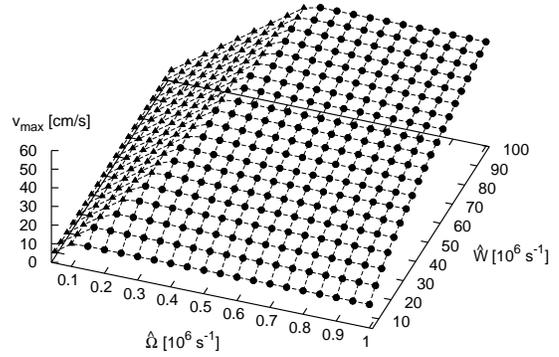}
\end{center}
\caption{\label{fig3} Limit $v_{max}$ for ``diodic'' behaviour,
$\epsilon = 0.01$;
three level atom, the mass is the mass of Neon,
$\Delta x = 15 \mum$, $x_S = 140 \mum$, $x_P = 170 \mum$, $x_W = 260 \mum$.}
\end{figure}

The results are shown in Fig. \ref{fig2}. In  
a velocity range, the ``diodic''
behaviour holds, i.e. $\fabsq{\hat{R} (v)} \approx 0$, $\fabsq{\hat{T} (v)}
\approx 1$
and $\fabsq{\hat{R} (-v)} \approx 1$, $\fabsq{\hat{T} (-v)} \approx 0$ ($v>0$).
In this range the other transmission and reflection
coefficients for incidence in the first state are zero.
The upper velocity boundary, $v_{upper}$,
for the diode with incidence from the left
is due to the breakdown of
the STIRAP effect \cite{kuklinski.1989} (A spontaneous 
decay rate $\Gamma$ from state 2 to state 1
does not alter $v_{upper}$ significantly  
for $\Omega/\Gamma\gtrsim 100$.)  
This boundary can be increased by increasing $\hat{\Omega}$.
The lower (negative) velocity boundary $v_{lower}$ for right incidence, 
due to the 
inability of the reflecting laser to block fast atoms,     
decreases when $\hat{W}$ increases,    
so that both boundaries
can be adjusted independently from each other. We may define  
$v_{max}>0$ as the minimum of 
$v_{upper}$ and $|v_{lower}|$.
More precisely, it is defined
by imposing that all scattering probabilities from the ground state be small 
except the ones that define the diode (i.e., the probability 
for transmission to $3$ from 
the left and for reflection to $1$ from the right),  
$\sum_{\alpha=1}^3 (|R_{\alpha 1}^l|^2+|T^r_{\alpha 1}|^2)
+\sum_{\alpha=1}^2
(|R^r_{\alpha+1,1}|^2+|T^l_{\alpha 1}|^2)+(1-|T_{31}^l|^2)+
(1-|R^r_{11}|^2)<\epsilon
$
for all $v_{min} \le v \le v_{max}$ with $v_{min} = 0.25\cms$.
In Fig. \ref{fig3}, $v_{max}$
is plotted versus $\hat{\Omega}$ and $\hat{W}$. For the intensities considered 
$v_{max}$ is in the ultracold regime below $1$ m/s. 
In the $v_{max}$ surface, $|v_{lower}|$ due to reflection failure 
is more restrictive in the hillside represented by circles, 
whereas 
$v_{upper}$, due to STIRAP failure, is more restrictive in the 
hillside with triangles. Considering the scales used for $\hat{\Omega}$ 
and $\hat{W}$, reflection failure is in general 
more problematic than STIRAP failure.


There is also a lower, positive-velocity boundary for the
STIRAP effect, i.e. the  
STIRAP effect breaks down at extremely low velocities, $0 < v \ll v_{min}$,
with the laser intensities (Rabi frequencies) of the numerical example.
This may appear contradictory since one expects 
better  adiabatic transfer at lower velocities. Indeed this is the case, but 
only as long  as the semiclassical approximation is valid for the translational 
motion. For sufficiently low velocities the quantum aspects of translational 
motion become important and atomic reflection occurs.    

%
\begin{figure}[t]
\begin{center}
\includegraphics[angle=0,width=0.3\linewidth]{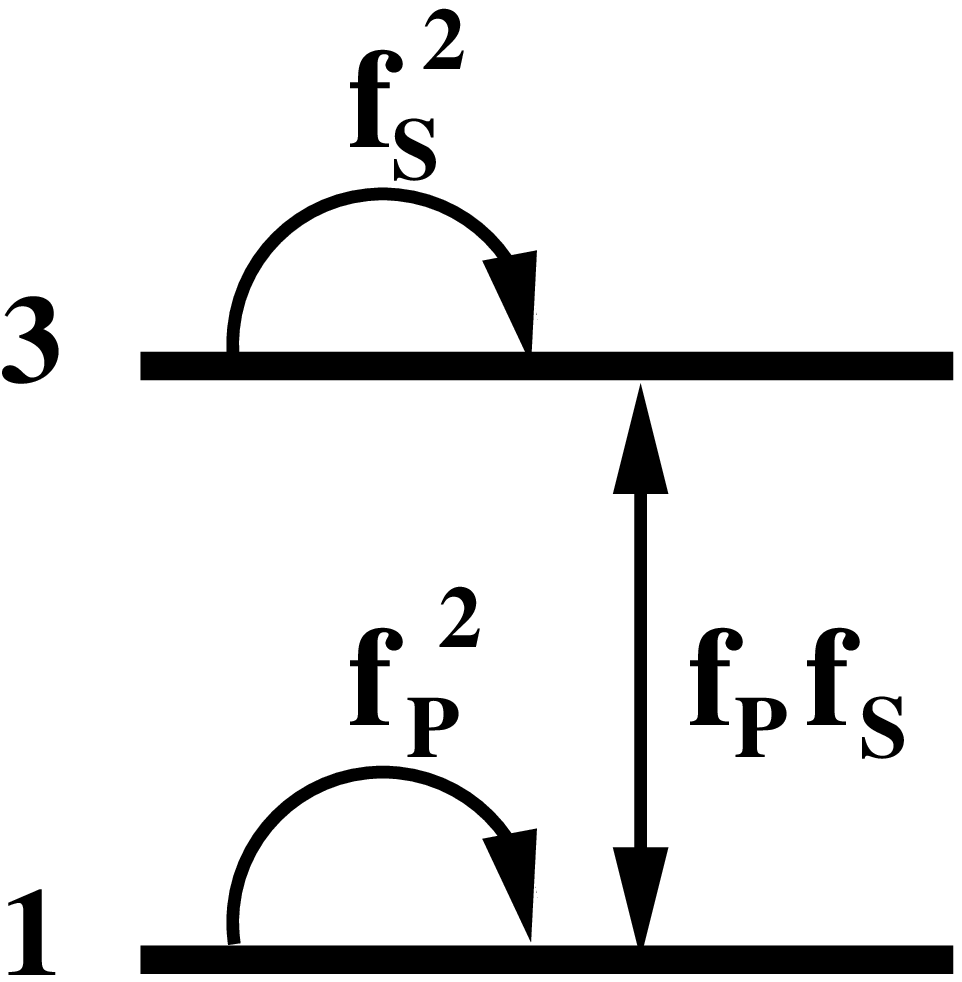}\hspace{0.5cm}
\includegraphics[angle=0,width=0.4\linewidth]{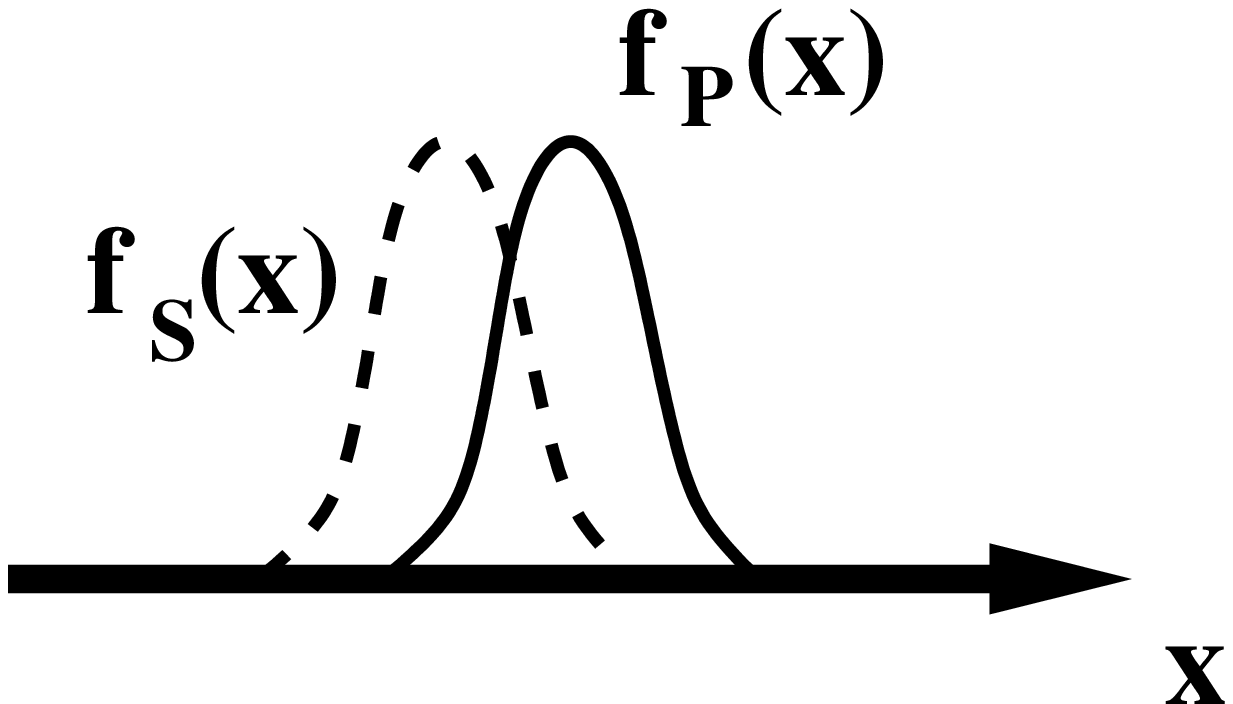}
\end{center}
\caption{\label{fig5}Schematic connection of the atom levels
by the different lasers (left figure) and the
order of the functions $f_S$, $f_P$ (right figure).}
\end{figure}
%
Notice that the diode behaviour can also be obtained for a
two level atom.  
It is well known \cite{carroll.1990}
that the three level Hamiltonian (\ref{ham}) with $W=0$
can be reformulated in the form of
a two-level one, but here we use a different idea to construct directly a
two-level potential with the ``diodic'' property.
Assume first 
that we can neglect the kinetic term
and that the motion in $x$ direction is classical. Let us define the
two position dependent eigenvectors of the two level potential $V'$ to be
\begin{eqnarray*}
\zeta_1 (x) &=& \frac{1}{\sqrt{f^2_P(x)+f^2_S(x)}}
\left(\begin{array}{c}f_S(x)\\-f_P(x)\end{array}\right),\\
\zeta_2 (x) &=& \frac{1}{\sqrt{f^2_P(x)+f^2_S(x)}}
\left(\begin{array}{c}f_P(x)\\f_S(x)\end{array}\right).
\end{eqnarray*}
With the order of $f_S,f_P \ge 0$ shown in Fig. \ref{fig5} we get for Gaussian 
(or similar) functions $f_S$ and $f_P$ the
asymptotic properties
\begin{eqnarray*}
\zeta_1(-\infty)=\left(\begin{array}{c}1\\0\end{array}\right)&,&
\zeta_1(+\infty)=\left(\begin{array}{c}0\\-1\end{array}\right),\\
\zeta_2(-\infty)=\left(\begin{array}{c}0\\1\end{array}\right)&,&
\zeta_2(+\infty)=\left(\begin{array}{c}1\\0\end{array}\right).
\end{eqnarray*}
This means that ground and excited
state are asymptotically swapped. $\zeta_1$ should correspond
to the eigenvalue $\lambda_1=0$ which results in adiabatic transfer from ground
to excited state if the atom impinges from the left, and 
$\zeta_2$ should correspond to
$\lambda_2=(\hbar/2)(f^2_P(x)+f^2_S(x)) \gg 0$,
so there will be nearly full reflection if the atom impinges
from the right. The eigenfunctions and eigenvalues 
define $V'$ and 
%
%
the two-level Hamiltonian is 
\begin{eqnarray}
H_{2L} = \frac{p_x^2}{2m} + \frac{\hbar}{2} \left(\begin{array}{cc}
f^2_P(x) & f_P(x) f_S(x)\\
f_P(x) f_S(x) & f^2_S(x)
\end{array}\right).
\label{ham2}
\end{eqnarray}
We have calculated the scattering amplitudes numerically
with $f_P (x) = \hat{f} \;\Pi(x, x_P)$ and 
$f_S (x) = \hat{f} \;\Pi(x, x_S)$ for right and left incidence
and observed the diodic behaviour, see 
Fig. \ref{fig2}.
The two-level Hamiltonian 
can be also used as a diode for incidence in the excited state. Then it
works in the opposite direction, i.e. 
$\fabsq{T^r_{13}(v)} \approx 1$, $\fabsq{R^r_{33}(v)} \approx 0$ and
$\fabsq{T^l_{31}(v)} \approx 0$, $\fabsq{R^l_{33}(v)} \approx 1$. This
is not the case for the Hamiltonian (\ref{ham}) unless an additional potential
acting on the third level is added.

\begin{figure}[t]
\begin{center}
\includegraphics[angle=0,width=\linewidth]{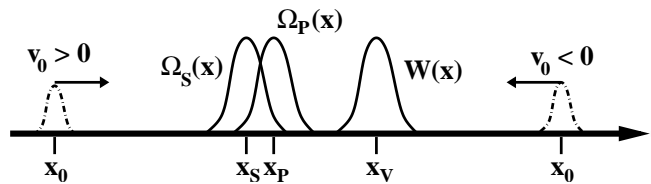}
\end{center}
\caption{\label{fig6} Scheme for the time-dependent
simulation including decay. The mass is the mass of Neon, 
$x_S = 140 \mum$, $\hat{\Omega}_S = 0.2 \Msi$,
$x_P = 170 \mum$, $\hat{\Omega}_P = 0.2 \Msi$, $x_W = 260 \mum$,
$\hat{W} = 10 \Msi$, $\Delta x = 15 \mum$,
$x_0 = 40 \mum$ ($v_0>0$) or $x_0 = 360 \mum$ ($v_0<0$),
and $\Delta v_0 = 0.1 \cms$.}
\end{figure}

Let us return to the three-level atom to study the possible effect of
decay 
from the third state to the first state with a relatively
small decay rate $\gamma$. This is unlikely a spontaneous process but it 
can be forced   
by a laser coupling of the third state to an 
auxiliary state decaying to the ground state. 
The process may be characterized by an effective decay rate 
from $3$ to $1$
\cite{ruschhaupt.2004}.
We examine the time-dependent case, see Fig. \ref{fig6}, by means of 
a one-dimensional master equation
which includes the effect of recoil (see \cite{hensinger.2001})
\begin{eqnarray}
\lefteqn{\frac{\partial}{\partial t} \rho
= - \frac{i}{\hbar} [H_{3L}, \rho]_-
- \frac{\gamma}{2} \{|3\ra \la 3|,\rho\}_+}\nonumber\\
&+& \gamma \int_{-1}^{1} \!\!du\; \frac{3}{8} (1+u^2) \;
\fexp{i\frac{mv_{rec}}{\hbar} u x}
\,|1\!>\nonumber\\
& & \times \, \la 3|\rho|3\ra \, \la 1|\,
\fexp{-i\frac{mv_{rec}}{\hbar} u x}.
\label{master_eq}
\end{eqnarray}
The initial state at $t=0$ is $\rho(0)=|\Psi_0\!><\!\Psi_0|$,
namely a Gaussian wave packet with mean velocity $v_0$, 
\begin{eqnarray*}
\Psi_0 (x) = \frac{1}{N} \left(\begin{array}{c} 1\\0\\0\end{array}\right)
\fexp{-\frac{\Delta v_0 m}{2\hbar} (x-x_0)^2 
+ i \frac{v_0m}{\hbar} x},
\end{eqnarray*}
where $N$ is a normalization constant.
%
%
%
\begin{figure}[t]
\begin{center}
\includegraphics[angle=-90,width=0.95\linewidth]{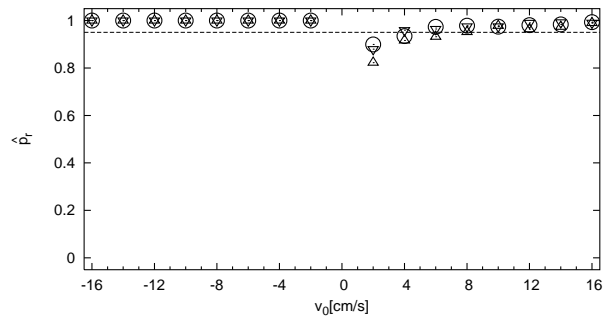}
\end{center}
\caption{\label{fig7} Probability $\hat{p}_r$ of traveling
to the right after $t_{max} = 400 \mum/v_0$;
$v_{rec} = 3 \cms$, $\gamma = 20 \si$ (down-pointing triangles);
$v_{rec} = 3 \cms$, $\gamma = 40 \si$ (up-pointing triangles);
$v_{rec} = 6 \cms$, $\gamma = 20 \si$ (circles);
$n=1000$ trajectories; the dashed line indicates $\hat{p}_r=0.95$;
other parameters in Fig. \ref{fig6}. 
}
\end{figure}
%
We solve the master equation 
by using the quantum jump technique
\cite{carmichael.book}.
Let $t_{max}$ be a sufficient large time such that the resulting wave packet
$\Psi_j (t_{max})$ of nearly every quantum ``trajectory'' $j$ separates
in right and left moving parts far from the interaction region but
possibly  with
third state components (not decayed yet at $t_{max}$).
By averaging over all trajectories we get 
\begin{eqnarray}
\hat{p}_r\!\!&=&\!\!\int_0^\infty \!\!\!\! dv\, (\la v|\rho_{11}(t_{max})|v\ra
 +\la v|\rho_{33}(t_{max})|v\ra)
\label{def_pr}
\end{eqnarray}
which is plotted in Fig. \ref{fig7} as a function of $v_0$ 
for different 
$\gamma$ and $v_{rec}$. The error bars, defined by the absolute
difference between averaging over $n/2$ and $n$ trajectories, are smaller 
than the symbol size. 

A value $\hat{p}_r (v_0) \approx 1$ for $v_0<0$ means
that nearly all atoms coming from the
right are reflected. 
The reflection probability
is not affected by the decay since the reflected 
atoms are rarely excited during the collision.

A value $\hat{p}_r (v_0) \approx 1$ for $v_0>0$ means
that nearly all atoms coming from the
left are transmitted and will be finally
in the ground state moving to the right.
This is true for $v_0 \ge 8 \cms$ (with $\hat{p}_r (v_0) \ge 0.95$)
for all examined combinations of decay rate $\gamma$ and recoil
velocity $v_{rec}$.
Therefore for not too low velocities a large part of the
atoms will be transmitted and stay finally in the ground state,
i.e. the atom diode works also with decay and recoil, with the advantage
that decay prevents the backward motion of excited atoms.   
The decrease of $\hat{p}_r$ for low, positive velocities is due to 
the atom decay
before passing the potential $W(x)\hbar/2$.

Summarizing, we have presented a simple model for an atom diode
that can be realized with laser interactions,  
a device which can be passed by the atom in one direction but not 
in the opposite direction.

\begin{acknowledgments}
We thank G. C. Hegerfeldt for commenting on the manuscript.
We acknowledge support by 
UPV-EHU (00039.310-13507/2001), ``Ministerio de
Ciencia y Tecnolog\'\i a''
and FEDER (BFM2003-01003).
AR acknowledges a fellowship within the Postdoc-Programme of the
German Academic Exchange
Service (DAAD).
\end{acknowledgments}
\end{document}